\documentclass[aps,amsfonts, twocolumn]{revtex4}

\usepackage{hyperref}
\usepackage{wasysym}
\usepackage{graphicx}
\usepackage{dsfont}
\usepackage{latexsym}
\usepackage{amssymb}

\def\<{\langle}
\def\>{\rangle}
\def\[{\left\lbrack}
\def\]{\right\rbrack}
\def\({\left(}
\def\){\right)}
\newcommand{\be}{\begin{equation}}
\newcommand{\ee}{\end{equation}}
\newcommand{\ea}{\end{eqnarray}}
\newcommand{\ba}{\begin{eqnarray}}

\newcommand{\vep}{{\varepsilon}}

\newcommand{\cl}{{\cal L}}

\newcommand{\ch}{{\cal H}}

\newcommand{\diag}{\mbox{diag}}

\newcommand{\bs}{\mbox{\hspace{-.1in}}}

\begin{document}

\title{Evolution of Anisotropies in   Eddington-Born-Infeld Cosmology}

\author{Davi C. Rodrigues}
\email{drodrigues@fis.puc.cl}
\affiliation{Departamento de F{\'\i}sica,
Universidad de Santiago de Chile, Casilla 307, Santiago, Chile \& \\ Departamento de F{\'\i}sica, P. Universidad Cat\'olica de Chile, Casilla 306, Santiago, Chile. }

\begin{abstract}
    Recently a Born-Infeld action for dark energy and dark matter  that uses additional affine connections was proposed.  At  background level, it was shown that the new proposal can mimic the standard cosmological evolution. In Bianchi cosmologies, contrary to the scalar field approach (e.g., Chaplygin gas), the new approach leads to anisotropic pressure, raising the issues of  stability of the isotropic solution under anisotropic perturbations and, being it stable, how the anisotropies evolve. In this work, the Eddington-Born-Infeld proposal  is extended to a Bianchi type I scenario and residual post-inflationary anisotropies are shown to decay in time. Moreover,  it is shown that the shears decay following a damped oscillatory  pattern, instead of the standard exponential-like decay.    Allowing for some fine tuning on the initial conditions,  standard theoretical bounds on the shears can be avoided.  \\[0.1in] 
\small{ Keywords: Dark energy, dark matter, Bianchi cosmology, isotropization.}
\end{abstract}

\maketitle

\section{Introduction}

One of the greatest  puzzles in current fundamental physics is the nature of dark energy and dark matter \cite{revdark, revdarkm}. Observationally their existence is well rooted in the results of many independent experiments,  e.g. \cite{exp, wmap5cosmo}, while theoretically the understanding of these entities is far from satisfactory.  The simplest way to model the dark sector, i.e. dark energy and dark matter, in good agreement with observations is through the now standard $\Lambda$CDM model \cite{wmap5cosmo}. Following this approach, at background level, dark matter enters as an additional dust contribution and dark energy as a positive cosmological constant. 

Considering their apparent dissimilarities, that dark matter and dark energy might be two faces of a single mysterious entity is an appealing proposal. This direction was first explored through the employment of the Chaplygin gas and its generalized version\cite{chap}, which use a (generalized) Born-Infeld action for scalar fields. Albeit  compatible with with cosmological data  at background level  \cite{chapback}, once density perturbations are considered the  Chaplygin gas approach has no satisfactory answer  to  the structure formation and the cosmic microwave background power spectrum \cite{chappert} . 

Recently, motivated from a general relativity extension to spaces whose space-time metric can be degenerated, a new model that describes an unified dark sector was proposed \cite{bimax}. In this Eddington-Born-Infeld model (or Einstein-Eddington-Born-Infeld), the space-time metric interacts with the Eddington action \cite{eddrev} through a Born-Infeld coupling. 

The Eddington action was proposed as an alternative action for gravitation \cite{eddrev}. Its fundamental fields are the affine connections and no metric appears in the action. It reads
\be
	S_{\mbox{\tiny Edd}}[\Gamma] = \int \sqrt{| R_{\mu \nu}(\Gamma)|} d^4x.
	\label{eddaction}
\ee
In above, $\Gamma^\rho_{\alpha \beta} = \Gamma^\rho_{\beta \alpha }$ are  independent affine connections and $| R_{\mu \nu} |$ is the modulus of the determinant of the symmetric  Riemann curvature tensor. The action (\ref{eddaction}) does not depend on the space-time metric but it is equivalent to the Einstein-Hilbert action in the presence of a non-null cosmological constant \cite{edddual}. Considering that current cosmological observations favors a non-null cosmological constant, the latter point is welcome. Nevertheless,  the coupling of gravity in this formulation to the rest of the universe is not a straightforward issue. The fundamental fields are the affine connections, but direct couplings of these to other structures are typically inconsistent, since the former are not tensors \footnote{Another possibility is to use the Riemann tensor in place of the metric. Considering the coupling between gravity and electromagnetism, this nonstandard coupling form was explored in \cite{eddEM}.}.

The above property  turns to be a welcome feature in the Eddington-Born-Infeld model. In the latter, the additional affine connections that come from the Eddington part are interpreted as the dark sector fundamental fields, which couples to the rest of the universe only through gravity.

Besides the aforementioned motivation, the employment of additional connections in cosmology, instead of scalar fields, is a new curious possibility which deserves  to be further investigated. The Eddington-Born-Infeld model can be seen as an affine connection version for the Chaplygin gas approach, where the scalar kinetic term inside the determinant is replaced by the curvature tensor for the additional ``dark"  connections.
	
	The purpose of this work is to analyze the Eddington-Born-Infeld  cosmology beyond the isotropic and homogeneous background.  Of particular interest is post-inflationary shear perturbations, which will be studied in  a Bianchi I framework. It generalizes the flat Friedmann-Lema\^{\i}tre-Roberton-Walker cosmology to homogeneous and anisotropic spaces without rotation. For reviews on Bianchi cosmologies see  \cite{bianchirev}. For some recent developments on the Bianchi I phenomenology see \cite{bianchiinflation1, bianchiinflation2} and references therein. While homogeneous scalar fields in Bianchi backgrounds exert isotropic pressure, preserving the shears tendency to decay, the approach with additional affine connections generate anisotropic pressure for backgrounds not exactly isotropic. Hence, {\it a priori}, the latter approach could   lead to an  anisotropic universe in conflict with observations\cite{bianchirev, shearsbound}, even if the universe anisotropy just after inflation was negligible. In the following sections it is shown that this {\it a priori } drawback for using additional connections in cosmology is not present in the Eddington-Born-Infeld model, the shears also decay in this model (although not monotonically).  In particular this implies that the isotropic solution found in \cite{bimax} is stable under shear perturbations. Moreover, with the purpose of disclosing model predictions, the way the shears decay is also analyzed, and it is shown that  it follows a characteristic pattern.

In the following section we review the Eddington-Born-Infeld model, establish notation and introduce some procedures that will be useful for the following sections.  Sec. III is devoted to the study general properties of the model, for both late times and early times,  that do not rely on the initial conditions that parametrize the anisotropies evolution. Sec IV presents numerical solutions for the Eddington-Born-Infeld model in a Bianchi I background, it  illustrates   Sec. III results,  shows the evolution of anisotropies at small redshifts --- a region that the Sec. III approach is inconclusive --- and presents two peculiar solutions that can avoid standard theoretical bounds on the shears \cite{bianchirev, shearsbound}. Finally, in Sec. V, our conclusions and perspectives are presented.

\section{Revisiting the Eddington-Born-Infeld Action and its Homogeneous and Isotropic Cosmology}

This section reviews some key results of \cite{bimax} and introduces procedures that will be useful to the following sections. 

The action proposed in \cite{bimax} reads
\ba
	\label{action}
	&& \bs \bs S[g,C, \Psi] =  \int  \cl_{m} (\Psi, g) ~d^{4}x + \\[.1in]
	&& 	+ \frac  k2  \int    \[ ~ \sqrt {|g_{\mu \nu}|}  R  + \frac 2 {\alpha l^{2}}  ~ \sqrt{  | g_{\mu \nu} - l^2 K_{\mu \nu}|}  ~\]  d^4x,  \nonumber
\ea
where $(g_{\mu \nu})$ is the space-time metric, $k \equiv 1 / (8 \pi G)$, $\Psi$ stands for any additional matter fields, $|X_{\mu \nu}|$ is the absolute value of the $(X_{\mu \nu})$ determinant,  $K_{\mu \nu}$ is the symmetric Ricci tensor constructed with the symmetric connection $C_{\mu \nu}^\rho$ and $R$ is the standard curvature scalar constructed with the space-time metric and the metric connection $\Gamma(g)$. The dimensions of the constants $l^2$ and $\alpha$ read  $[ l^{2} ] = m^{-2}$, $[\alpha] = 1$.  The affine connection $C^\rho_{\mu \nu}$ couples to  $g_{\mu \nu}$ through a Born-Infeld interaction, and  only influence additional fields indirectly through the coupling of $g_{\mu \nu}$ to them.

In the absence of $\cl_m$, the equations of motion of the action (\ref{action}) read
\ba
	\label{eomA1}
	 && \bs G_{\mu \nu}   =  \frac 1 {\alpha {l^2}} ~ \sqrt { \frac { |g - l^2 K|  }{|g| }}   ~ g_{\mu \alpha} (g - l^2 K)^{  \alpha \beta} g_{\beta \nu},  \\[.1in]
	\label{eomA2}
	 && \bs \nabla_\kappa^C \( \sqrt { |g - l^2 K|} (g - l^2 K)^{  \alpha \beta} \) = 0,
\ea
where the indices inside the determinants were omitted, $((g - l^2 K)^{  \alpha \beta})$ is the inverse of the matrix $(g_{  \alpha \beta} - l^2 K_{  \alpha \beta})$  and $\nabla^C$ is the covariant derivative constructed with the  $C$ connection.

The  equations of motion of the pure Eddington action (\ref{eddaction}) are given by (\ref{eomA2}) without $g_{\mu \nu}$. In order to express such equations in a form closer to the Einstein equations, one introduces an auxiliary non-degerate symmetric tensor $q^{\mu \nu}$ such that $ \nabla^C_\alpha q^{\mu \nu} = 0$, which implies that $C^\alpha_{\beta \rho}$ can be written as functions of $q_{\mu \nu}$ and derivatives. Letting $q_{\mu \nu} \propto K_{\mu \nu}$, Eq. (\ref{eomA2}) (without $g_{\mu \nu}$) becomes an identity and the equations of motion are now simply $q_{\mu \nu} \lambda =  K_{\mu \nu}(q)$, where $\lambda$ is a constant. Interpreting $q_{\mu \nu}$ as the space-time metric, one sees that  the latter equations are just  Einstein equations for a de Sitter space-time with $\lambda$ as the cosmological constant. For other details on the correspondence of this formulation with the standard one, see \cite{edddual}.

A similar trick is also useful for the action  (\ref{action}). The main differences are that the auxiliary tensor $q_{\mu \nu}$ is not the space-time metric, and that the value of the arbitrary constant $\lambda$ corresponds to a rescaling of $q_{\mu \nu}$ that has  no dynamical role. The resulting new expressions for the equations of motion are

\ba
	\label{G}
	&& G_{\mu \nu}   =  - \Lambda ~ \sqrt { \frac {|q|}{|g| }}   ~ g_{\mu \alpha} q^{  \alpha \beta} g_{\beta \nu},  \\[.1in]
	\label{K}
	 &&K_{\mu \nu}(q)  =  {\Lambda }  \[ \alpha q_{\mu \nu} + (  1 - \alpha) ~ g_{\mu \nu} \].
\ea
In above we have  re-parametrized the model from $\{l^2, \alpha\}$ to $\{\Lambda, \alpha\}$ using the same relation of \cite{bimax}, 
\be
	l^2 = \frac 1{1- \alpha} \frac 1 \Lambda, 
	\label{l2l}
\ee
but a different normalization for  $q_{\mu \nu}$ was selected. To change to the original normalization of \cite{bimax} one should replace every $q_{\mu \nu}$ by $( 1 - \alpha )~ q_{\mu \nu} $. 

The simplest nontrivial solution of Eqs. (\ref{G}, \ref{K}) are found for  $C_{\mu \nu}^\rho = \Gamma_{\mu \nu}^\rho $.  This implies that  $q_{\mu \nu} = \gamma g_{\mu \nu}$, for constant $\gamma$. From (\ref{G}), one concludes  that this corresponds to a de Sitter solution with $\Lambda \gamma$ as the cosmological constant; and requiring compatibility with (\ref{K}) leads to $\gamma = 1$. The latter defines the selected $q_{\mu \nu}$ normalization.

Eq. (\ref{K}) can be set in the standard form of Einstein equations,
\begin{equation}
	\label{Q}
	Q_{\mu \nu} = \Lambda (1 - \alpha) \( \frac {\alpha ~ q_{\mu \nu} }{ \alpha - 1} + g_{\mu \nu} - \frac 12 g_{\alpha \beta} q^{\alpha \beta} q_{\mu \nu} \),
\end{equation} 
where $ Q_{\mu \nu} \equiv K_{\mu \nu} - \frac 12 K_{\alpha \beta} q^{\alpha \beta} q_{\mu \nu}$. \\[.1in]

Now we turn to the isotropic and homogeneous cosmology spawned by the equations (\ref{G}, \ref{Q}) with
\ba
	\label{giso}
	&& \bs (g_{\mu \nu}) = \diag \pmatrix{-N^2 & a^2 & a^2 & a^2}, \\[.1in]
	\label{qiso}
	&& \bs (q_{\mu \nu})  = \diag  \pmatrix{-X^2 & Y^2 & Y^2& Y^2}.
\ea
 All the above variables depend on time alone. The equations of motion, in the presence of usual matter and radiation, can be stated as
\ba
	\label{g1i}
	&& 3 H^2 = N^2 \( \Lambda \frac {N Y^3}{X a^3} + \frac {\rho_b}k + \frac {\rho_r}k \), \\[.1in]
	\label{q1i}
	&& 3 H_Y^2 = \frac {\Lambda (1 - \alpha) X^2}{2 } \(  \frac {2 \alpha}{ 1- \alpha} - \frac {N^2}{X^2} + 3 \frac {a^2}{Y^2} \), \\[.1in]
	\label{emc}
	&& 3 a^2 H  \frac {X Y}N  = \frac d {dt} \(\frac {N Y^3}X \),
\ea
where $H \equiv \dot a / a$, $H_Y \equiv \dot Y / Y$, $\rho_b$ is the baryon energy density and $\rho_r$ is the radiation energy density. Once the temporal gauge is fixed, one has three equations for three unknowns. The above equations are the $G_{00}$ (\ref{g1i}) constraint, the $Q_{00}$  constraint (\ref{q1i}),  and the energy momentum conservation for the ``dark fluid" (\ref{emc}). The latter can be found  from the Bianchi identities of either $G_{\mu \nu}$ or $Q_{\mu \nu}$ together with (\ref{G}) or (\ref{Q}). 

Eqs.(\ref{g1i}, \ref{q1i}, \ref{emc}) were the equations selected in \cite{bimax} for evaluating the numerical solutions, which is the simplest choice in the isotropic picture. Nevertheless, for plotting anisotropic cosmologies one does not have the possibility of selecting only the first order equations. To prepare to the next section, we will use the second order equations,
\ba
	\label{g2i}
	&& \bs 2 \dot H + 3 H^2 - 2 H H_N  =   \Lambda \frac {NXY}a - \frac {N^2 \rho_r}{3k},  \\[.1in]
	&& \bs 2 \dot H_Y + 3 H_Y^2 - 2 H_Y H_X  =  \nonumber \\[.1in]
	\label{q2i}
	&& =  \frac {\Lambda ( 1- \alpha) X^2}{2 } \(  \frac {2 \alpha} {1-\alpha} - \frac 1 {X^2} + \frac {a^2}{Y^2} \),
\ea
together with (\ref{emc}) as the equations to be solved numerically. With $H_X \equiv \dot X / X$ and  $H_N \equiv \dot N / N$. Eqs. (\ref{g1i}, \ref{q1i})  constrains the initial conditions. \\[.1in]

In order to plot the numerical solutions, we fix the temporal gauge with $N = 1$ and search for a consistent set of initial conditions. We will use $X_0$, $H_0$, $Y_0$, $\dot Y_0$ to denote the values of $X$, $H$, $Y$, $\dot Y$ at present time $t_0$, and we set  $a(t_0) = 1$. The initial conditions will be set at $t = t_0$. Letting (\ref{g1i}) to set the value of  $X_0$ as a function of $H_0$ and $Y_0$, the $\dot Y$ initial condition reads
\ba
\dot Y_0 = && \bs   \pm \frac{  \sqrt{1-\alpha} ~ \Lambda^{\frac 32} ~ Y^4_0} {\sqrt{6}   \left(-3  H_0^2+ \frac{\rho_b}k+ \frac {\rho_r}k \right)} \[ \frac{2 \alpha}{1-\alpha}+\frac{3}{Y^2_0}- \right. \nonumber \\[.1in]
	\label{dotY0}
	&& \left .  - \frac{  \left(-3  H_0^2+ \frac {\rho _b}k + \frac{\rho _r}k \right)^2}{  \Lambda^2 ~ Y^6_0} \]^\frac 12. 
\ea

\begin{figure}[htbp]
	\includegraphics[width=0.49\textwidth]{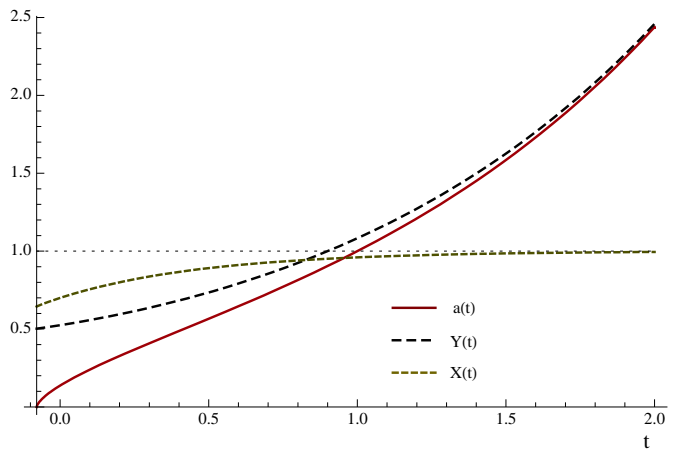}
	\caption{Plot for all the Eddington-Born-Infeld functions, with $\Omega_b = 0.05$,   $\Omega_r = 8.49 \times 10^{-5}$, $\Omega_\Lambda = 0.72$, $H_0 = 0.98$, $t_0 = 1$ and  $\alpha = 0.99$. The values of $Y_0$,  $\dot Y_0$ and $X_0$ are inferred from the latter.  The big bang happens at $t=-0.787...$.  This deviation from zero is just outside the one sigma level error for the universe age \cite{wmap5cosmo}. It is  a negligible deviation, in particular since this model parameters have yet to be properly fit directly from the cosmological data\cite{ebicmb, ebirot, ebinext}.}
	\label{graphisodS}
\end{figure}

\begin{figure}[htbp]
	\includegraphics[width=0.49\textwidth]{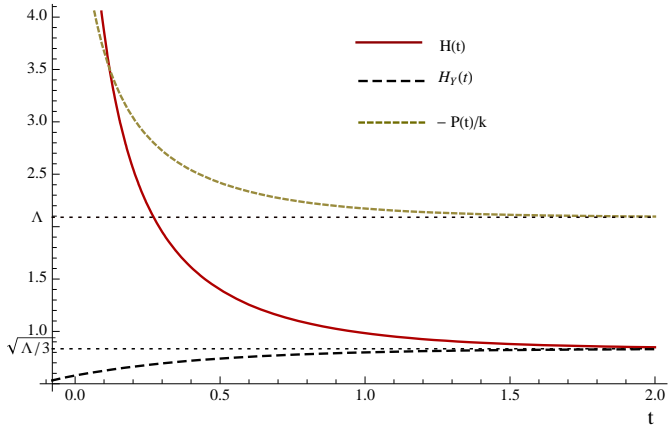}
	\caption{Evolution of $H = \dot a /a$,  $H_Y = \dot Y /Y$ and $P/k =  -  \Lambda  {XY}/a$,  considering the same parameters of Fig.(\ref{graphisodS}). All these three quantities are  in above considered as adimensional quantities (using  $t_0 =1$). Although the ratio of the pressure $P$ by the energy density goes to zero as $t$ approaches  the big bang  \cite{bimax}, the pressure $P$ does not go to zero.}
	\label{isoHHYP}
\end{figure}

Besides $\rho_r$ and $\rho_b$, two parameters, $\alpha$ and $\Lambda$, and two initial conditions, $Y_0$ and $H_0$, are still free.  The proper way of fixing them is to search for the best fit directly from cosmological data \cite{ebicmb, ebirot, ebinext}. Here we  follow the original proposal  and assume that this best fit should lead to values of $\rho_r$, $\rho_b$, $\Lambda$ and $H_0$ close to the ones found from the $\Lambda$CDM best fit. Any choice of these parameters close to the values found from WMAP5+BAO+SN \cite{wmap5cosmo} is perfectly consistent with this paper proposal, and we will take precisely those values --- except for radiation, for it a mean value is adopted. Regarding the values of $Y_0$ and $\alpha$,  $\alpha$ is set to $0.99$ (it should be close to one to allow solutions that are close to the standard cosmological model \cite{bimax}), and $Y_0$ is set such that the Hubble parameter directly evolves toward  $\sqrt{\Lambda/3}$, leading to $Y_0 = 1.083$ \footnote{In \cite{bimax} the approach for fixing $Y_0$ was finding the closest scale factor evolution to the standard cosmological model, leading to $Y_0 = 1.059$. But the analytical analysis is more transparent with $Y_0 = 1.083$, in particular since $H$ goes  directly to $\sqrt{\Lambda /3}$, instead of going to a lower value and than slowly increasing to $\sqrt{\Lambda /3}$ for $t > 1$.  There are no changes in our conclusions if either of  these  values of $Y_0$ is used.}. Fig. (\ref{graphisodS}) shows a plot for the evolution of all the isotropic Eddington-Born-Infeld functions. And Fig.(\ref{isoHHYP}) shows the evolution of  $H$, $H_Y$ and the pressure induced in (\ref{g2i}) by the new fluid, it will be useful for the next section.

\section{Anisotropic perturbations}

The approach of this section is to study the behavior of the anisotropies assuming that they are fluctuations  around the isotropic solution. The isotropic quantities, whose analytical solutions are unknown, are analyzed numerically, while the anisotropic quantities are analyzed analytically. In this model, as it will be shown in the following section, there are six initial conditions that parametrize the evolution of the anisotropies, therefore a pure numerical analysis for some sets of initial conditions does not unveil significantly the general picture. In this section  it will be shown that, independently on the anisotropic initial conditions, the isotropic solution acts as an atractor for both early and late times, and that the Eddington-Born-Infeld isotropization has a characteristic pattern.\\[.1in]

The Bianchi I cosmology follows from general relativity with the homogeneous metric \cite{bianchirev} 
\be
	\label{gbianchi}
	(g_{\mu \nu}) = \diag \pmatrix{-N^2 & a_1^2 & a_2^2 & a_3^2}.
\ee

	Introducing $H_i \equiv \dot a_i / a_i$, its four equations of motion can be expressed as \footnote{We are using a notation as close  as possible to the standard one in isotropic cosmologies. There is no need to proceed with the 1+3 covariant decomposition here.} 
\ba
	\label{bianchi1}
	&& \bs H_1 H_2 + H_2 H_3 + H_1 H_3 = \frac {N^2}k ~ \rho, \\[.1in]
	\label{bianchi2}
	&&  \bs  \bs \bs   \sum_{ \matrix{j=1  \cr  j \not = i } }^3 \( \dot H_j + H_j^2 - H_N H_j\)+ \frac {H_1 H_2 H_3}{H_i} = - \frac {N^2} k p_i.
\ea
These generalize the  usual Friedmann equations for a flat anisotropic universe without rotation, with energy density $\rho$ and pressures  $p_i$. 

	Using ${\cal H} \equiv \frac 13 \sum_j H_j$, the shears evolution is found from
\be
	\label{sheare}
	\frac d {dt} \( H_i - H_k \) = (H_N - 3 {\cal H}) \( H_i - H_k \) + \frac {N^2} k \(p_i - p_k \). \\[.2in]
\ee

The cosmology proposed in \cite{bimax} is generalized to a Bianchi type I one by using the metric (\ref{gbianchi}) and the auxiliary tensor field
\be
	(q_{\mu \nu} )= \diag \pmatrix{-X^2 & Y^2_1 & Y^2_2 & Y_3^2}.
\ee

The equations of motion (\ref{G}, \ref{Q})  can be written in the form of  (\ref{bianchi1}, \ref{bianchi2}) with the appropriate metric variables, energy density and pressures. The energy density and pressures  induced by the $q_{\mu \nu}$ field on the equations for the $g_{\mu \nu}$  dynamics read, henceforth fixing $N=1$, 
\ba
	&& \bs \rho_q =  k \Lambda  \frac {Y_1 Y_2 Y_3}{X a_1 a_2 a_3}, \\[.1in]
	&& \bs {p_q}_i  = - k \Lambda  \frac {X Y_1 Y_2 Y_3}{a_1 a_2 a_3} \frac{a_i^2}{Y_i^2}.
\ea
Analogously, the metric induces a fluid on the $q_{\mu \nu}$ equations whose energy density and pressures  are
\ba
	&& \bs \bs \bs  \rho_g = \frac { k \Lambda (1-\alpha)} {2 } \(  \frac {2  \alpha }{ 1 - \alpha} - \frac 1 {X^2} + \sum_{j=1}^3 \frac{a_j^2}{Y_j^2} \), \\[.1in]
	&& \bs \bs \bs {p_g}_i = - \frac { k \Lambda (1 - \alpha)} {2 }  \(  \frac {2 \alpha} {1-\alpha}  + \frac 1 {X^2} + \sum_{j=1}^3 \frac{a_j^2}{Y_j^2} - 2\frac{a_i^2}{Y_i^2}  \).
\ea
In the above we could have introduced another constant different from  ${ k}$, but that is superfluous since  the additional $C_{\alpha \beta}^\rho$ connections do not couple to anything aside from gravity. The above energies and pressures reduce to their corresponding isotropic quantities  if $a_i = a$ and $Y_i = Y$ (see \ref{g1i}, \ref{q1i}, \ref{g2i}, \ref{q2i}).

With the purpose of separating the isotropic contribution from the anisotropic ones, we set
\ba
	&& {p_q}_i  = {\cal P} + {\xi}_i, \\[.1in]
	&& H_i = {\cal H} + \vep_i \\[.1in]
	&& {H_Y}_i ={\cal H}_Y + \gamma_i.
\ea
After the gauge is fixed  ($N=1$) a residual symmetry is still present, namely the rescalings of the spacial coordinates by constant factors, leading to the following transformations for the scale factors: $a_i \rightarrow c_i ~ a_i$ and $Y_i \rightarrow c_i ~ Y_i$, where the $c_i$'s are arbitrary non-null constants. One should take care when dealing with isotropization directly from the metric components; in particular,  that all the metric components approach the same value is not sufficient (nor necessary) for isotropization to happen; it can be just an artifact of the selected coordinate system. On the other hand, the quantities $H_i$, $ {H_Y}_i$ and ${p_q}_i$ are all invariant under the above symmetry, and the six independent quantities $ {p_q}_i - {p_q}_j$, $H_i - H_j$, $ {H_Y}_i - {H_Y}_j$   fully characterize the anisotropies.

To first order on the anisotropic perturbations ($\vep_i, \xi_i, \gamma_i$), the equations of motion satisfied by  $\ch, \ch_Y$ and $\cal P$ are the same of  their corresponding isotropic quantities ($H$, $H_Y$, $p_q$) if we set   $\sum_i  \xi_i = \sum_i  \vep_i = \sum_i  \gamma_i = 0$.  This result only depends on the assumption that $\vep_i, \gamma_i$ are small in regard to $\ch, \ch_Y$ respectively. That is, $\vep_i$ is assumed to be small in regard to $\ch$, but its relation to other  quantities is arbitrary.

The evolution of the pressure anisotropies (anisotropic stresses) read
\ba
	&&  \bs \bs \frac d {dt } ( p_{q_i} -p_{q_j})  =  \frac d {dt } ( \xi_{i} -\xi_{j})  = \( {\ch}_Y - {\ch} + H_X \) ( \xi_i - \xi_j) + \nonumber    \\[.1in] 
	\label{pe}
	&&  \bs \bs + 2 {\cal P} ( \vep_i - \gamma_i - \vep_j + \gamma_j ) + 2 \xi_i (\vep_i - \gamma_i) - 2 \xi_j (\vep_j - \gamma_j). 
\ea
Whenever the anisotropies are small in regard to their corresponding isotropic values, the two last terms from  (\ref{pe}) can be dropped. 

All the dynamics of the anisotropies can be found from the two versions of  (\ref{sheare}) (one for $g_{\mu \nu}$ and the other for $q_{\mu \nu}$) and (\ref{pe}).  In a standard Bianchi I cosmology there are two groups of two interacting anisotropic quantities, which are the shears ($H_i - H_j$) and the pressure anisotropies ($p_i - p_j$). In the Eddington-Born-Infeld extension to a Bianchi I scenario, the evolution of the latter depends on the physical shears ($H_i - H_j$) and the shears from $q_{\mu\nu}$ (${H_Y}_i - {H_Y}_j$). There is not an independent second set of pressure anisotropies, see
 (\ref{pqpg}).

\vspace{.2in}
\subsection{Early time behavior}

To some extent,  how the shears in the Eddington-Born-Infeld model will behave just after inflation depends on the role of the connections $C^\rho_{\alpha \beta}$ during inflation. However, at early times, if the anisotropic quantities are smaller than their isotropic counterparts, or the shears will decay or their derivatives will approach   zero. There is no guarantee that the shears will decay from the early universe  to the  present time, and Figs. (\ref{scmbearly}, \ref{scmblate}) illustrate that, but the isotropic solution acts as an attractor. 

For the shears modulus to increase it is necessary that $|\xi_i - \xi_j| >  | 3 k \ch (\vep_i - \vep_j)|$. The assumption of small anisotropies does not rule out this possibility at early times, for the pressure exerted by the unified dark fluid starts at minus infinity. Moreover, no upper limit in $|\xi_i - \xi_j|$ can avoid that inequality, since there is no lower bound for the shears. Nevertheless, if that inequality holds at early times, then 
\be
	\label{pewithoutvep}
	\frac d {dt} (\xi_i - \xi_k)  \sim (\ch_Y - \ch + H_X) (\xi_i - \xi_j) + 2 {\cal P} (- \gamma_i + \gamma_k).
\ee
The latter comes from
\ba
	&& \bs \bs \bs |(\ch_Y - \ch + H_X) (\xi_i - \xi_j)| >  \\[.1in]
	&& \bs  |(\ch_Y - \ch + H_X)  3 k {\cal H} (\vep_i - \vep_j)| \gg 2| {\cal P} (\vep_i - \vep_j)|, \nonumber
\ea
since, as $t$ goes to zero, $ { \cal H} $ increases faster than $|\cal P|$ (see also Fig.(\ref{isoHHYP})) --- the first evolves proportionally to $a^{-3/2}$ (matter phase) or $a^{-2}$ (radiation phase) while the second to $a^{-1}$.

Therefore, if after inflation the pressure anisotropies have a relevant contribution to the shears evolution, the former will quickly decay if the last term of  (\ref{pewithoutvep}) can be neglected. It might happen that the pressure anisotropies will not decrease just after inflation, for this to happen it is necessary that  $2| {\cal P} (- \gamma_i + \gamma_k)| >  |(\ch_Y - \ch + H_X) (\xi_i - \xi_j)|$.  Which implies, at early times, that  $|( H_X - 3 \ch_Y) ( \gamma_i - \gamma_k)| \gg |\Lambda(1-\alpha)/\rho_q ~ (\xi_i - \xi_j)|$, and hence, from (\ref{sheare}, \ref{pqpg}), that  $| \gamma_i - \gamma_k|$ will decay.

In conclusion, it is possible that the modulus of the physical shears increase at early times, but if this happens the pressure anisotropies will decrease, reducing the shears derivative. The isotropic solution acts as an attractor at early times.

\vspace{.1in}
\subsection{Late time behavior}
For the late time behavior, another approach is capable of unveiling more details.   Using the following relations that  come directly from the definitions,
\ba
	\label{pqpg}
	&& {p_q}_i - {p_q}_j = - \frac {X^2 ~ \rho_q  } { k \Lambda(1-\alpha)}     ( {p_g}_i - {p_g}_j ), \\[.1in]
	&&  \dot {\cal P} =  ( \ch_Y - \ch + H_X ) {\cal P}, \\[.1in]
	&& \dot \rho_q =   ( 3 (\ch_Y - \ch)  - H_X)\rho_q,
\ea
it is possible to write the second order expressions for the evolution of the shears and pressure anisotropies as follows,
\ba
	&& \bs  \bs \frac {d^2} {dt^2} (\vep_i - \vep_j) = \( - 3 \dot \ch + 2 \frac {\cal P}{k} \) (\vep_i - \vep_j) - 3 \ch \frac d {dt} (\vep_i - \vep_j)  \nonumber \\[.2in]
	\label{o1}
	&& + \frac 1k ( \ch_Y - \ch + H_X) (\xi_i - \xi_j) - 2 \frac {\cal P} k (\gamma_i - \gamma_j), 
\ea	

\ba	
	&& \bs \bs \frac {d^2} {dt^2} (\xi_i - \xi_j) = \( \dot \ch_Y - \dot \ch + \dot H_X + 2 \frac {\cal P} k - 2 \frac {{\cal P} \Lambda(1-\alpha)}{\rho_q } \) (\xi_i - \xi_j) \nonumber \\[.2in]
	&& +  (  \ch_Y -  \ch +  H_X) \frac d {dt} (\xi_i - \xi_j) + 2 {\cal P} (- \ch_Y + \ch) (\gamma_i - \gamma_j) \nonumber \\[.2in]
	\label{o2}
	&& + 2 {\cal P} (\ch_Y - 4 \ch + H_X)(\vep_i - \vep_j),
\ea

\ba
	&& \bs \bs \frac {d^2} {dt^2} (\gamma_i - \gamma_j)  = \( \dot H_X - 3 \dot \ch_Y + 2 \frac{{\cal P}\Lambda(1-\alpha)}{\rho_q } \) (\gamma_i - \gamma_j) \nonumber \\[.2in]
	&& + (H_X - 3 \ch_Y) \frac d {dt} (\gamma_i - \gamma_j) - 2 \frac {{\cal P}\Lambda(1-\alpha)}{\rho_q } (\vep_i - \vep_j) \nonumber \\[.2in]
	\label{o3}
	&& + \frac {2 \Lambda(1-\alpha)}{ \rho_q} (  \ch_Y - \ch - H_X) (\xi_i - \xi_j).
 \ea

Albeit less compact, the above form is useful for interpreting the evolution of anisotropies as the interaction of three parametric damped oscillators. 

The differential equation $ \ddot x =  - b \dot x - \omega^2 x  $ describes a damped oscillator with damping factor $b >0$ and resonance frequency $\omega$. The effective frequency of oscillation is  $ \frac 12 \sqrt{4 \omega^2 - b^2}$. If this value is complex the $x$  evolution is given by hyperbolic trigonometric functions, otherwise oscillations appear for $x$. We will use these observations as approximations to the late time behavior of  (\ref{o1},\ref{o2},\ref{o3}); firstly without interaction, i.e. neglecting the two last terms in each of the above equations \footnote{In this sense, without interactions is not a physical limit,  it is just a first approach to the Eqs. (\ref{o1}, \ref{o2}, \ref{o3}).}.

\begin{figure}[htbp]
	\includegraphics[width=0.5\textwidth]{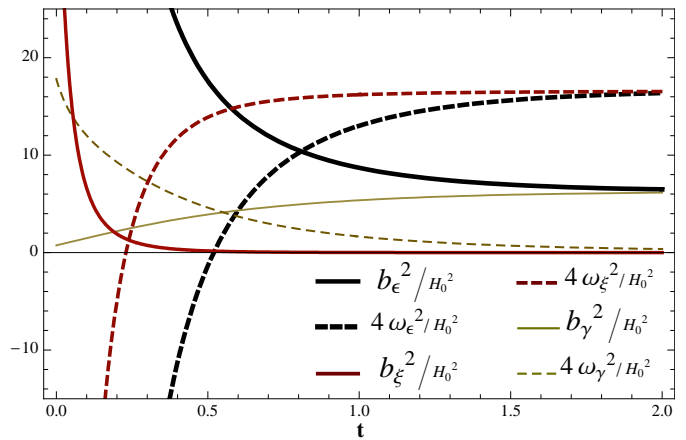}
	\caption{Evolution of the damping and resonance parameters on time. Generated from the same isotropic solution displayed  in Fig.(\ref{graphisodS}). In particular, even though at late times this model isotropic version behaves as a standard de Sitter universe, its Bianchi I version does not display the typical exponential decay for the shears, instead they decay oscillating (see Conclusions for other details).}
	\label{b2c}
\end{figure}

From the numerical solutions of  the three damping coefficients,  $b_\vep = 3 \ch, b_\gamma = - (H_X - 3 \ch_Y) $ and $ b_\xi = - (  \ch_Y -  \ch +  H_X)$, it is found that they are always positive, leading to a tendency towards isotropization. In Fig. (\ref{b2c}) the values of $b^2_\vep, b^2_\gamma$ and $b^2_\xi$ are plotted and compared to  $4 \omega^2_\vep = 4 \(  3 \dot \ch - 2 \frac {\cal P}{k} \)$, $4 \omega^2_\gamma =  4 \(- \dot H_X + 3 \dot \ch_Y - 2 \frac{{\cal P} \Lambda (1- \alpha) }{\rho_q } \)$ and $ 4 \omega^2_\xi = 4 \( -\dot \ch_Y + \dot \ch - \dot H_X - 2 \frac {\cal P} k + 2 \frac {{\cal P} \Lambda (1- \alpha)}{\rho_q } \) $.

The effective frequency of oscillation $ \frac 12 \sqrt{4 \omega^2 - b^2}$  for the physical shears and the pressures anisotropies  becomes real after $ t \simÊ0.8 ~ t_0$ and $ t \sim 0.2~ t_0$ respectively. These quantities oscillate with approximate asymptotic effective (angular) frequencies of $ 2 / t_0$ and $3/(2 t_0)$, and hence their half periods are  $3/2 ~ t_0$ and $ 2  ~ t_0$  respectively. These late time low frequency oscillations   can reduce dramatically the decrement of the physical shears, when compared to the standard Bianchi I exponential decay, but do not spoil the isotropization tendency, since the damping factors are always positive. 

The shears from $q_{\mu \nu}$ have an intrinsic oscillatory tendency at high redshift, but with low and quickly decreasing frequency. While for late times, it simply decay with an exponential-like behavior.

When the two last terms of (\ref{o1}, \ref{o2}, \ref{o3}) are considered, turning on the interaction among the three damped parametric oscillators, the picture at $t \lesssim 1$ can change considerably depending on the initial conditions (see next section).  During these times, due to the interactions, some oscillations with ill defined large period ($ \gtrsim 0.5$) can appear. High frequency oscillations cannot appear since all of the oscillators have low frequency. For latter times, the behavior is less dependent on the initial conditions.   The physical shears, in particular,  oscillate with a well defined low frequency and with decreasing amplitude, interacting only with the shears from $q_{\mu \nu}$ (since $\ch_Y - \ch + H_X \rightarrow 0$), whose natural tendency is to decay in an exponential way.

\section{Numerical solutions for the evolution of anisotropies}

In this section we present an approach to the numerical analysis of this problem and numerical solutions that complement Sec. III results. For late times ($t>1$) the anisotropies always decay oscillating with well defined low frequency, but many possibilities are left for the details on the anisotropies evolution at both high and low redshift. This section  displays some peculiar possibilities of this model, considering  small anisotropies that fluctuate around the isotropic solution presented in Sec. II. 

The Eddington-Born-Infeld model with a Bianchi I background is described by seven equations of motion and seven unknowns,  once the temporal gauge is fixed. Analogously to the isotropic case as presented in Sec. II, the equations  selected to be solved numerically are the six second order ones (i.e., (\ref{bianchi2}) with the appropriate variables) and  energy momentum conservation,
\be
	 X Y_1 Y_2 Y_3 \(\sum_{j=1}^3  \frac {a_j^2}{Y_j^2} H_{j} \) = \frac d {dt} \( \frac {Y_1 Y_2 Y_3}{X} \).
\ee

We use $X_0, {H_0}_i, {Y_0}_i$ and ${\dot Y}_{0_i}$ to denote the values of  $X, {H}_i, {Y}_i$ and ${\dot Y}_i$ at the present time $t_0$, and set $a_i(t_0) = 1$. The first order equation  (\ref{bianchi1}) for the metric determines  $X_0$ as a function of ${H_0}_i, {Y_0}_i$, while the first order equation for $q_{\mu \nu}$ determines one of the $\dot {Y_0}_i$'s a function of ${H_0}_i, {Y_0}_i$ and the two others $\dot {Y_0}_i$'s.

In general, the value of $X_0$ in the anisotropic case is different from the isotropic one.  It remains to be imposed that the isotropic quantities $H_0$ and $Y_0$ are the mean values of the anisotropic quantities. To this end, one sets $H_0 = \frac 13 \sum_{i=1}^3 {H_0}_i$ and $Y_0^3 = {Y_0}_1 {Y_0}_2{Y_0}_3$. The latter, for small anisotropies, are sufficient for finding a $X_0$ that is equal to the isotropic one and ${H_Y}_0 =\frac 13   \sum_{i=1}^3 {{H_Y}_0}_i$.

\begin{figure}[htbp]
	\includegraphics[width=0.5\textwidth]{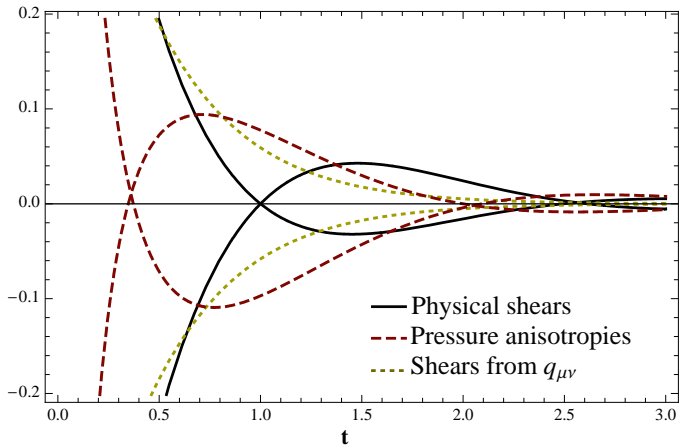}
	\caption{Evolution of all anisotropic quantities at late time. All the quantities in this plot refer to the corresponding adimensional ratios, i.e. $(H_i - H_j)/ \ch$, $({H_Y}_i - {H_Y}_j)/ \ch_Y$, $({p_q}_i - {p_q}_j)/ {\cal P}$. Only the two shears with greater magnitude are displayed, together with the pressure anisotropies and $q_{\mu \nu}$ shears of same indices. All the anisotropies were set to be small at $t=1$, but with a considerable magnitude difference. The physical shears were set to be $~ 10^{-5}$, being close to the usual bounds, and the others were set to  $~ 10^{-1}$. This solution only has small anisotropies at late times.}
	\label{numsol3}
\end{figure}

\begin{figure}[htbp]
	\includegraphics[width=0.5\textwidth]{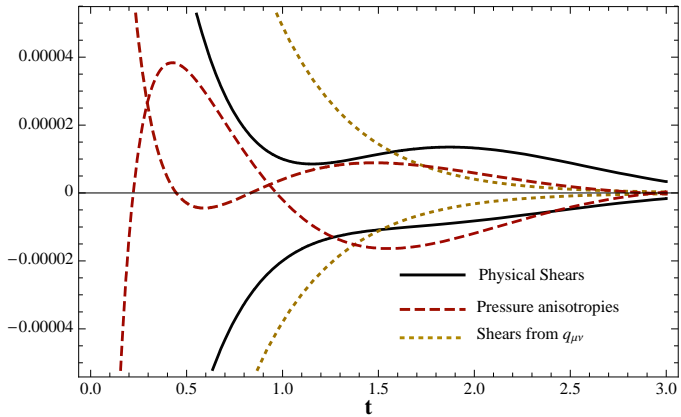}
	\caption{Like in Fig.(\ref{numsol3}), but all anisotropies are set to have about the same order $\sim 10^{-5}$ at $t_0$, and the sign of the pressure anisotropies at $t_0$ was inverted, preventing the shears to continue to decay before reaching zero.}
	\label{numsol10a}
\end{figure}

Figs. (\ref{numsol3}, \ref{numsol10a}) exemplify the evolution of anisotropies at late times for two sets of initial conditions set at $t_0$. Some details on the evolution at low redshift depend strongly on the initial conditions (e.g., the shears may or may not become null at $t \lesssim 1$), but at latter times their behavior follows the general pattern described in Sec. III. 

\begin{figure}[htbp]
	\includegraphics[width=0.5\textwidth]{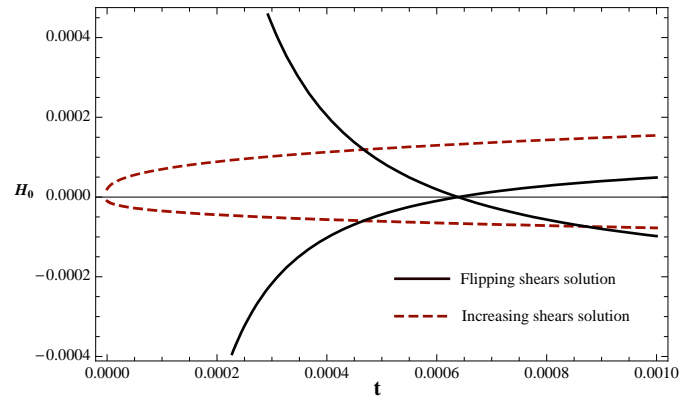}
	\caption{Evolution of two sets of shears with initial conditions set at decoupling time (about $10^{-5} t_0$ after the big bang). In above the instant $t=0$ was shifted to be the decoupling time and the shears are measured in $H_0$ unities. For the dashed solution it was used, at decoupling: $H_i - H_j \sim 10^{-9} \ch$, ${H_Y}_i - {H_Y}_j \sim 10^{-2} \ch_Y$, $ {p_q}_i - {p_q}_j \sim 10^{-3} {\cal P}$. For the solid solution, the same as above, but  $H_i - H_j \sim 10^{-5} \ch$.  }
	\label{scmbearly}
\end{figure}

\begin{figure}[htbp]
	\includegraphics[width=0.5\textwidth]{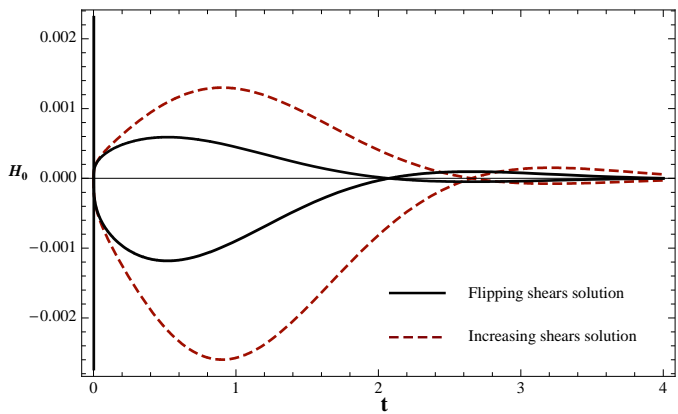}
	\caption{The late time behavior of the same solutions of Fig.(\ref{scmbearly}).}
	\label{scmblate}
\end{figure}

From the numerical solution of the isotropic case (or from the anisotropic equations solution with isotropic initial conditions), one can infer the value of all model parameters at any time, and hence generate numerical solutions from initial conditions at any time. Figs. (\ref{scmbearly}, \ref{scmblate}) show two sets of solutions for the shears evolution with nonstandard features. Both were generated from initial conditions set at  matter-radiation decoupling time ---  which is a natural choice since the strongest shear observational constraints come from the cosmic microwave background (CMB) \cite{bianchirev}.  The dashed solution shows that, in spite of the isotropic solution being an attractor, it is possible that the shears have been increasing since the last scattering surface formation and up to the present time. This idea was explored recently in other models as a way for solving certain cosmological anomalies, e.g. \cite{anisdark1, anisdark3, anisdark5}.  The solid shears solution shows the possibility that the shears could have been high at at the last scattering an high at present time, but with inverted signs. This possibility can  significantly reduce the standard observational constraints on the shears ---  at the price of adding one coincidence problem in cosmology, since the CMB will seem more isotropic at current time than at the past or the future. Note that all solutions of this model have shears that invert sign, since at late times the shears approach zero like damped oscillators. The special feature of this solution (the solid one)  is when the first sign flip happens.

\section{Conclusions}

The Eddington-Born-Infeld (EBI) model \cite{bimax} is a new candidate for an unified dark sector and uses additional connections, instead of scalar fields, to model dark energy and dark matter. The approach with additional connections makes its extension to Bianchi cosmologies nontrivial since it induces anisotropic pressures (stresses).

In this work the Eddington-Born-Infeld cosmology was extended to a Bianchi type I cosmology and it was shown that $i$) the full dynamics of anisotropies depends on two noninteracting groups of three anisotropic quantities (the physical shears, the shears from $q_{\mu \nu }$, and the pressure anisotropies);  $ii$) for both early and late times  the isotropic solution acts as an attractor, implying that the shears tend to decay, and that observational constraints on the shears can be fulfilled \cite{bianchirev, shearsbound}; $iii$) the shears do not decay monotonically, but they typically oscillate with decreasing amplitude; $iv$) letting some fine tuning on the initial conditions that parametrize this model anisotropies, we could construct two solutions qualitatively different from usual anisotropic models Figs. (\ref{scmbearly}, \ref{scmblate}); specifically,  these are capable of avoiding standard theoretical bounds on the shears.

Since here we have evaluated anisotropic fluctuations around the isotropic EBI solution found in \cite{bimax}, some of our results rely on the same assumption considered in \cite{bimax}, namely that some of the EBI isotropic parameters should have values similar to the corresponding ones from $\Lambda$CDM \cite{wmap5cosmo}. We stress that further EBI developments may lead to different parameters, for the Ref. \cite{wmap5cosmo} results were found assuming $\Lambda$CDM. On the other hand, many qualitative aspects of the Bianchi I EBI cosmology here presented do not depend on the precise values of those isotropic parameters; in particular, from  Eqs.(\ref{sheare},\ref{pe}) or (\ref{o1},\ref{o2},\ref{o3}) one sees the shears oscillations is a general feature of this cosmology\footnote{Shears oscillations, albeit following a different pattern, were also found in another cosmological context \cite{os}.}. There is yet much to explore in anisotropic EBI systems, we should return to them in a future work.

Collins and Hawking \cite{collinshawking} analyzed the  issue of stability of the Friedmann-Lema\^{\i}tre-Robertson-Walker (FLRW) cosmology under anisotropic perturbations, and found that the universe is unstable at late times in some Bianchi models. However, a cosmological constant at late times was not considered in their analysis, which is known to be sufficient to lead to stability \cite{wald}. At late times, the isotropic  EBI model  effectively behaves like FLRW with cosmological constant. Nevertheless, the anisotropic perturbation  induces a modification in the effective cosmological constant and it starts to exert anisotropic pressure. It can be inferred from Eq. (\ref{pe}) and Fig. (\ref{b2c}) that, if a shear perturbation occurs at late time, the shears will induce anisotropic pressure which will prevent the exponential decay\footnote{Actually the anisotropic pressure will at first help the shears to decay even faster than the standard exponential decay.}, but the shears will decay following a damped oscillatory behavior; hence  EBI cosmology is also stable at late times. 

Recently anisotropic forms of dark energy have been receiving increasing attention \cite{anisdark1}\cite{anisdark2}\cite{anisdark3}\cite{anisdark4}\cite{anisdark5}. These proposals, besides constraining what dark energy can and cannot be, are generating their first observational consequences and perhaps are the explanation behind some anomalies found in the standard cosmological model. The Eddington-Born-Infeld model naturally generates an anisotropic form of dark energy that is considerably intricate. It is not the purpose of this paper to pinpoint a set of initial conditions capable of solving a particular anomaly; but we would like to stress that standard bounds on the shears need not to be satisfied in this model. 

The present work is a step toward the proof that  the EBI model is in conformity with observational data for modeling dark energy and dark matter. Of special relevance to this proof is the  analysis of density fluctuations and consequences for the CMB anisotropies \cite{ebicmb}, since this  step the Chaplygin gas approach could not overcome as a  unified dark sector model \cite{chappert}. If the EBI model can indeed be a realistic ``quartessence" model it has yet  to be proven, but this paper results do not depend on that.

Considering the original motivation \cite{bimax}, the EBI model should be more than a phenomenological model for dark energy and dark matter; the role of the $C^\alpha_{\beta \rho}$ connections during inflation have yet to be evaluated and other theoretical issues are being evaluated.

\vspace{.2in}
\noindent
{ \bf Acknowledgments}

The author would like to thank M\'aximo Ba\~nados for suggesting the study of the anisotropic version of the EBI cosmology, and for useful discussions. The author also thanks Martin Makler for comments on the Chaplygin gas and Jorge Zanelli for a discussion on the Eddington action. This work was supported by FONDECYT-Chile grant n. 3070008.


\begin{thebibliography}{99}

\bibitem{revdark} V.~Sahni and A.~A.~Starobinsky, ``The Case for a Positive Cosmological Lambda-term,''  Int.\ J.\ Mod.\ Phys.\  D {\bf 9}, 373 (2000) [arXiv:astro-ph/9904398]; S.~M.~Carroll,``The cosmological constant,''  Living Rev.\ Rel.\  {\bf 4}, 1 (2001) [arXiv:astro-ph/0004075];  L.~Perivolaropoulos, ``Accelerating universe: Observational status and theoretical implications,''  AIP Conf.\ Proc.\  {\bf 848}, 698 (2006)  [arXiv:astro-ph/0601014]; E.~J.~Copeland, M.~Sami and S.~Tsujikawa,``Dynamics of dark energy,''  Int.\ J.\ Mod.\ Phys.\  D {\bf 15}, 1753 (2006)  [arXiv:hep-th/0603057]; M.~Ishak,``Remarks on the formulation of the cosmological constant/dark energy problems,''  Found.\ Phys.\  {\bf 37}, 1470 (2007)  [arXiv:astro-ph/0504416];  J.~P.~Uzan, ``The acceleration of the universe and the physics behind it,''  Gen.\ Rel.\ Grav.\  {\bf 39}, 307 (2007)  [arXiv:astro-ph/0605313]; T.~Padmanabhan, ``Dark Energy and Gravity,''  Gen.\ Rel.\ Grav.\  {\bf 40}, 529 (2008) [arXiv:0705.2533 [gr-qc]];    M.~S.~Turner and D.~Huterer,``Cosmic Acceleration, Dark Energy and Fundamental Physics,''  arXiv:0706.2186 [astro-ph].

\bibitem{revdarkm}  S.~Dodelson,``Modern Cosmology,'' {\it  Amsterdam, Netherlands: Academic Pr. (2003) 440 p}; V.~Sahni,``Dark matter and dark energy,''  Lect.\ Notes Phys.\  {\bf 653}, 141 (2004)  [arXiv:astro-ph/0403324]; G.~Bertone, D.~Hooper and J.~Silk,``Particle dark matter: Evidence, candidates and constraints,''  Phys.\ Rept.\  {\bf 405}, 279 (2005)  [arXiv:hep-ph/0404175].

\bibitem{exp}  A.~G.~Riess {\it et al.}  [Supernova Search Team Collaboration], ``Observational Evidence from Supernovae for an Accelerating Universe and a Cosmological Constant,''  Astron.\ J.\  {\bf 116}, 1009 (1998)  [arXiv:astro-ph/9805201]; D.~Clowe, M.~Bradac, A.~H.~Gonzalez, M.~Markevitch, S.~W.~Randall, C.~Jones and D.~Zaritsky,``A direct empirical proof of the existence of dark matter,'' Astrophys.\ J.\  {\bf 648}, L109 (2006)
  [arXiv:astro-ph/0608407];  D.~N.~Spergel {\it et al.}  [WMAP Collaboration],``Wilkinson Microwave Anisotropy Probe (WMAP) three year results: Implications for cosmology,''  Astrophys.\ J.\ Suppl.\  {\bf 170}, 377 (2007)  [arXiv:astro-ph/0603449]; M.~Tegmark {\it et al.}  [SDSS Collaboration],``Cosmological parameters from SDSS and WMAP,''  Phys.\ Rev.\  D {\bf 69}, 103501 (2004)  [arXiv:astro-ph/0310723].

\bibitem{wmap5cosmo} E.~Komatsu {\it et al.}  [WMAP Collaboration],``Five-Year Wilkinson Microwave Anisotropy Probe (WMAP) Observations:Cosmological Interpretation,''  arXiv:0803.0547 [astro-ph].


\bibitem{chap}  A.~Y.~Kamenshchik, U.~Moschella and V.~Pasquier,``An alternative to quintessence,''  Phys.\ Lett.\  B {\bf 511}, 265 (2001)
  [arXiv:gr-qc/0103004]; N.~Bilic, G.~B.~Tupper and R.~D.~Viollier, ``Unification of dark matter and dark energy: The inhomogeneous Chaplygin
  gas,''  Phys.\ Lett.\  B {\bf 535}, 17 (2002)  [arXiv:astro-ph/0111325]; M.~C.~Bento, O.~Bertolami and A.~A.~Sen,``Generalized Chaplygin gas, accelerated expansion and dark energy-matter unification,''  Phys.\ Rev.\  D {\bf 66}, 043507 (2002)  [arXiv:gr-qc/0202064]. 
    
\bibitem{chapback} T.~Padmanabhan and T.~R.~Choudhury, ``Can the clustered dark matter and the smooth dark energy arise from the  same scalar field?,''  Phys.\ Rev.\  D {\bf 66}, 081301 (2002)  [arXiv:hep-th/0205055].   M.~Makler, S.~Quinet de Oliveira and I.~Waga, ``Observational constraints on Chaplygin quartessence: Background  results,''  Phys.\ Rev.\  D {\bf 68}, 123521 (2003)  [arXiv:astro-ph/0306507];  ``Constraints on the generalized Chaplygin gas from supernovae  observations,''  Phys.\ Lett.\  B {\bf 555}, 1 (2003)  [arXiv:astro-ph/0209486]; R.~J.~Colistete, J.~C.~Fabris, S.~V.~B.~Goncalves and P.~E.~de Souza,``Bayesian Analysis of the Chaplygin Gas and Cosmological Constant Models using the SNe Ia Data,''
  Int.\ J.\ Mod.\ Phys.\  D {\bf 13}, 669 (2004)  [arXiv:astro-ph/0303338].
  
\bibitem{chappert}   H.~Sandvik, M.~Tegmark, M.~Zaldarriaga and I.~Waga, ``The end of unified dark matter?,''  Phys.\ Rev.\  D {\bf 69}, 123524 (2004)
  [arXiv:astro-ph/0212114];  L.~Amendola, F.~Finelli, C.~Burigana and D.~Carturan, ``WMAP and the Generalized Chaplygin Gas,''  JCAP {\bf 0307}, 005 (2003)  [arXiv:astro-ph/0304325]; R.~R.~R.~Reis, I.~Waga, M.~O.~Calvao and S.~E.~Joras, ``Entropy perturbations in quartessence Chaplygin models,''  Phys.\ Rev.\  D {\bf 68}, 061302 (2003)  [arXiv:astro-ph/0306004]; V.~Gorini, A.~Y.~Kamenshchik, U.~Moschella, O.~F.~Piattella and A.~A.~Starobinsky,
  ``Gauge-invariant analysis of perturbations in Chaplygin gas unified models  of dark matter and dark energy,''  JCAP {\bf 0802}, 016 (2008)
  [arXiv:0711.4242 [astro-ph]].




\bibitem{bimax} M. Ba\~nados, ``Eddington-Born-Infeld action for dark matter and dark energy",  Phys. Rev. D {\bf 77}, 123534 (2008) [arXiv:0801.4103 [hep-th]];  ``Eddington-Born-Infeld action and the dark side of general relativity,''  arXiv:0807.5088 [gr-qc].


\bibitem{eddrev} A.S. Eddington, ``The mathematical theory of Relativity", Cambridge University Press (1924); E. Schroedinger, ``Space-time structure", Cambridge University Press (1950). 

\bibitem{edddual} M.~Ferraris and J.~Kijowski, ``On The Equivalence Of The Relativistic Theories Of Gravitation,''  Gen.\ Rel.\ Grav.\  {\bf 14}, 165 (1982); E.~S.~Fradkin and A.~A.~Tseytlin, ``Quantum Equivalence Of Dual Field Theories,''  Annals Phys.\  {\bf 162}, 31 (1985).


\bibitem{eddEM} M.~Ferraris and J.~Kijowski, ``General Relativity Is A Gauge Type Theory,''   Lett.\ Math.\ Phys.\  {\bf 5}, 127 (1981);  N.~J.~Poplawski,
  ``The affine theory of gravitation and electromagnetism. II,''  arXiv:gr-qc/0701176;   ``On the Maxwell Lagrangian in the purely affine gravity,''  Int.\ J.\ Mod.\ Phys.\  A {\bf 23}, 567 (2008)  [arXiv:gr-qc/0702129].



\bibitem{bianchirev} G.~F.~R.~Ellis, ``The Bianchi models: Then and now,''  Gen.\ Rel.\ Grav.\  {\bf 38}, 1003 (2006);  J.~Wainwright,  G.F.R.~ Ellis,  (eds.): ``Dynamical Systems in Cosmology". Cambridge University Press (1997). 

\bibitem{bianchiinflation1} A.~E.~Gumrukcuoglu, C.~R.~Contaldi and M.~Peloso, ``Inflationary perturbations in anisotropic backgrounds and their imprint on
  the CMB,''  JCAP {\bf 0711}, 005 (2007)  [arXiv:0707.4179 [astro-ph]].
  
\bibitem{bianchiinflation2}   C.~Pitrou, T.~S.~Pereira and J.~P.~Uzan, ``Predictions from an anisotropic inflationary era,''  JCAP {\bf 0804}, 004 (2008)  [arXiv:0801.3596 [astro-ph]].

\bibitem{shearsbound}  J.~D.~Barrow, ``Cosmological limits on slightly skew stresses,''  Phys.\ Rev.\  D {\bf 55}, 7451 (1997)  [arXiv:gr-qc/9701038];  J.~D.~Barrow and R.~Maartens, ``Anisotropic stresses in inhomogeneous universes,''  Phys.\ Rev.\  D {\bf 59}, 043502 (1999)  [arXiv:astro-ph/9808268];  J.~D.~Barrow, R.~Maartens and C.~G.~Tsagas, ``Cosmology with inhomogeneous magnetic fields,''  Phys.\ Rept.\  {\bf 449}, 131 (2007)
  [arXiv:astro-ph/0611537].

\bibitem{ebicmb} M. Ba\~nados, P. Ferreira, C. Skordis, work in preparation.

\bibitem{ebirot} A. Reisenegger and N. Rojas,  work in preparation.

\bibitem{ebinext} M. Ba\~nados, M. Bennet, work in preparation.

\bibitem{os}   J.~D.~Barrow and J.~J.~Levin, ``Chaos in the Einstein-Yang-Mills equations,''  Phys.\ Rev.\ Lett.\  {\bf 80}, 656 (1998)  [arXiv:gr-qc/9706065]; J.~D.~Barrow, Y.~Jin and K.~i.~Maeda, ``Cosmological co-evolution of Yang-Mills fields and perfect fluids,''  Phys.\ Rev.\  D {\bf 72}, 103512 (2005)  [arXiv:gr-qc/0509097].


\bibitem{collinshawking}   C.~B.~Collins and S.~W.~Hawking, ``Why is the Universe isotropic?,''  Astrophys.\ J.\  {\bf 180}, 317 (1973).

\bibitem{wald} R.~W.~Wald,``Asymptotic behavior of homogeneous cosmological models in the presence of a positive cosmological constant,''  Phys.\ Rev.\  D {\bf 28}, 2118 (1983).

\bibitem{anisdark1}  T.~Koivisto and D.~F.~Mota, ``Dark energy anisotropic stress and large scale structure formation,''  Phys.\ Rev.\  D {\bf 73}, 083502 (2006)  [arXiv:astro-ph/0512135];   ``Accelerating Cosmologies with an Anisotropic Equation of State,'' Astrophys. J. {\bf 679}:1 (2008) [arXiv:0707.0279 [astro-ph]]; ``Anisotropic Dark Energy: Dynamics of Background and Perturbations,''  arXiv:0801.3676 [astro-ph]; ``Vector Field Models of Inflation and Dark Energy,''  arXiv:0805.4229 [astro-ph].

\bibitem{anisdark2} R.~A.~Battye and A.~Moss, ``Anisotropic perturbations due to dark energy,''  Phys.\ Rev.\  D {\bf 74}, 041301 (2006)  [arXiv:astro-ph/0602377].

\bibitem{anisdark3}J.~Beltran Jimenez and A.~L.~Maroto, ``Cosmology with moving dark energy and the CMB quadrupole,''  Phys.\ Rev.\  D {\bf 76}, 023003 (2007) [arXiv:astro-ph/0703483].


\bibitem{anisdark4}  D.~F.~Mota, J.~R.~Kristiansen, T.~Koivisto and N.~E.~Groeneboom,``Constraining Dark Energy Anisotropic Stress,''  arXiv:0708.0830 [astro-ph].

\bibitem{anisdark5}  D.~C.~Rodrigues,  ``Anisotropic Cosmological Constant and the CMB Quadrupole Anomaly,''
  Phys.\ Rev.\  D {\bf 77}, 023534 (2008)  [arXiv:0708.1168 [astro-ph]].



\end{thebibliography}
\end{document}